\documentclass[a4paper,12pt]{article}

\usepackage[dvips]{graphicx}
\usepackage{epsfig}

\begin{document}

\author{C. Barrab\`es\thanks{E-mail : barrabes@lmpt.univ-tours.fr}\\
\small Laboratoire de Math\'ematiques et Physique Th\'eorique,\\
\small CNRS/UMR 6083, Universit\'e F. Rabelais, 37200 TOURS,
France
\\\small and \\P. A. Hogan\thanks{E-mail : peter.hogan@ucd.ie}\\
\small School of Physics,\\ \small University College Dublin,
Belfield, Dublin 4, Ireland}

\title{On Generating Gravity Waves with Matter and Electromagnetic Waves}
\date{}
\maketitle

\begin{abstract}If a homogeneous plane
light--like shell collides head--on with a homogeneous plane
electromagnetic shock wave having a step--function profile then no
backscattered gravitational waves are produced. We demonstrate, by
explicit calculation, that if the matter is accompanied by a
homogeneous plane electromagnetic shock wave with a step--function
profile then backscattered gravitational waves appear after the
collision.
\end{abstract}
\thispagestyle{empty}
\newpage
\setcounter{equation}{0}
\section{Introduction}\indent

This paper is concerned with the production of gravitational waves
from the interaction of matter with light in the context of
Einstein--Maxwell classical field theory. We present an explicit
physical mechanism for the generation of gravitational waves by
such a process.

In general matter interacting with light does not produce
gravitational waves. A simple illustration of this is the head--on
collision of a homogeneous plane light--like shell with a
homogeneous plane electromagnetic shock wave having a profile
described by the Heaviside step--function (see \cite{BH1}
(eqs.(3.9)--(3.11)) where the electromagnetic shock wave is also
accompanied by a homogeneous plane light--like shell with no
resulting backscattered gravitational waves after the collision).
This solution of the vacuum Einstein--Maxwell field equations fits
the pattern of all known post--collision space--times (see for
example \cite{BH2}, \cite{SKMHH}): The line--element of the
space--time after the head--on collision of homogeneous, linearly
polarized, plane light--like signals has the Rosen--Szekeres form
\cite{R}\cite{S}
\begin{equation}\label{1.1}
ds^2=-e^{-U}\,(e^V\,dx^2+e^{-V}\,dy^2)+2\,e^{-M}\,du\,dv\
,\end{equation}with $U, V, M$ functions of $(u, v)$. The Maxwell
field has only two (real--valued) Newman--Penrose components $\phi
_0$ and $\phi _2$ which are both functions of $(u, v)$. Writing
$\hat\phi _0=e^{-U/2}\phi _0$ and $\hat\phi _2=e^{-U/2}\phi _2$,
the vacuum Maxwell field equations read
\begin{equation}\label{1.2}
\frac{\partial\hat\phi _0}{\partial u}=-\frac{1}{2}V_v\,\hat\phi
_2\ ,\qquad \frac{\partial\hat\phi _2}{\partial
v}=-\frac{1}{2}V_u\,\hat\phi _0\ ,\end{equation}with subscripts
denoting partial derivatives where convenient. Einstein's field
equations with the electromagnetic field here as source read
\cite{SKMHH}:
\begin{eqnarray}\label{1.3}
U_{uv}&=&U_u\,U_v\ ,\\
2\,U_{uu}&=&U_u^2+V_u^2-2\,U_u\,M_u+4\,\phi _2^2\ ,\\
2\,U_{vv}&=&U_v^2+V_v^2-2\,U_v\,M_v+4\,\phi _0^2\ ,\\
2\,V_{uv}&=&U_u\,V_v+U_v\,V_u+4\,\phi _0\,\phi _2\ ,\\
2\,M_{uv}&=&V_u\,V_v-U_u\,U_v\ .\end{eqnarray}Consider now the
head--on collision of a plane light--like shell of matter labelled
by a real parameter $k$ (in the sense that if $k=0$ then the
light--like shell is removed) with an electromagnetic shock wave
with amplitude $b$ and having a Heaviside step--function profile.
We take the history of the signal labelled by $k$ to be found in
the region $u>0,\ v<0$ of the space--time with line--element (1.1)
while the history of the signal labelled by $b$ is found in the
region $u<0,\ v>0$ of the space--time (the region $u<0,\ v<0$ is
taken to be flat so that the signals are non--interacting before
collision). To solve the field equations above in the region
$u>0,\ v>0$ of the space--time after the collision requires the
following conditions on the boundary of this region of
space--time: for $u>0, v=0$ we require
\begin{equation}\label{1.4}
e^{-U}=(1-k\,u)^2\ ,\ V=0\ ,\ M=0\ ,\ \phi _2=0\
,\end{equation}and for $v>0, u=0$ we require
\begin{equation}\label{1.5}
e^{-U}=\frac{1}{1+b^2v^2}\ ,\ V=0\ ,\ e^M=1+b^2v^2\ ,\ \phi
_0=\frac{b}{1+b^2v^2}\ .\end{equation}The solution is given by
\cite{BH1}
\begin{eqnarray}\label{1.6}
e^{-U}&=&(1-k\,u)^2+\frac{1}{1+b^2v^2}-1\ ,\\
e^{-M-\frac{1}{2}U}&=&\frac{1-k\,u}{(1+b^2v^2)^{3/2}}\ ,\\
\hat\phi _0&=&\frac{b}{(1+b^2v^2)^{3/2}}\ ,\end{eqnarray} together
with $V=0$ and $\phi _2=0$. The Newman--Penrose components of the
Weyl tensor calculated with the metric tensor given via (1.1) are
\begin{eqnarray}\label{1.7}
\Psi _0&=&-\frac{1}{2}(V_{vv}-U_v\,V_v+M_v\,V_v)\ ,\\
\Psi _1&=&0\ ,\\
\Psi _2&=&\frac{1}{4}(V_u\,V_v-U_u\,U_v)\ ,\\
\Psi _3&=&0\ ,\\
\Psi _4&=&-\frac{1}{2}(V_{uu}-U_u\,V_u+M_u\,V_u)\
.\end{eqnarray}Clearly all of these components with the exception
of $\Psi _2$ vanish for the solution given immediately above
(because $V=0$). This is a Petrov Type D Weyl tensor and since
$\Psi _0$ and $\Psi _4$ vanish there is no backscattered
gravitational radiation present. \emph{The purpose of this paper
is to establish the existence of backscattered gravitational waves
if the light--like shell here is accompanied by an electromagnetic
shock wave}. To achieve this it is sufficient to consider a small
amplitude accompanying electromagnetic shock wave as a
perturbation of the space--time with the functions $U, M, \phi _0$
given by (1.10)--(1.12) with $V=0$ and $\phi _2=0$. Without making
the small amplitude assumption the collision problem posed here
appears intractable (it is a non--trivial problem even with the
small amplitude assumption, as will appear below) notwithstanding
the fact that from a physical point of view it is very clear and
simple.

\setcounter{equation}{0}
\section{The Perturbed Space-Time}\indent

If the light--like shell above is accompanied by a homogeneous
plane electromagnetic shock wave with a step--function profile and
amplitude labelled by $a$ then the boundary conditions (1.8) must
be replaced by: for $u>0, v=0$ we require
\begin{equation}\label{2.1}
e^{-U}=\frac{(1-k\,u)^2}{1+a^2u^2}\ ,\ V=0\ ,\ e^M=1+a^2u^2\ ,\
\phi _2=\frac{a}{1+a^2u^2}\ .\end{equation}Taking the amplitude
$a$ to be small we shall neglect squares and higher powers of $a$
and thus replace (2.1) by: for $u>0, v=0$ we require
\begin{equation}\label{2.2}
e^{-U}=(1-k\,u)^2\ ,\ V=0\ ,\ M=0\ ,\ \phi _2=a\
.\end{equation}The field equations and the boundary conditions
(1.9) and (2.2) give the following, which are useful later (in
particular for determining the boundary conditions to be satisfied
by the function ${\cal K}$ in (2.19)): for $u>0, v=0$:
\begin{equation}\label{2.2'}
V_u=0\ ,\ V_v=\frac{2\,a\,b\,u}{1-k\,u}\ ,\ \phi
_0=\frac{b}{1-k\,u}\ ,\end{equation} and for $v>0, u=0$:
\begin{equation}\label{2.2''}
V_u=2\,a\,b\,v\ ,\ V_v=0\ ,\ \phi _2=a\ .\end{equation}

It is convenient to write $V$ in the form
\begin{equation}\label{2.3}
V=\log\left (\frac{1+A}{1-A}\right )\ ,\end{equation}with $A(u,
v)$ small of order $a$ (which we write as $A=O(a)$). Thus we have
\begin{equation}\label{2.4}
V_u=2\,A_u=O(a)\ ,\qquad V_v=2\,A_v=O(a)\ .\end{equation}We assume
in addition that $\phi _2=O(a)$. Neglecting $O(a^2)$--terms we
have henceforth
\begin{equation}\label{2.5}
e^{-U}=(1-k\,u)^2+\frac{1}{1+b^2v^2}-1\ .\end{equation}Equations
(1.3) and (1.7) give, neglecting $O(a^2)$--terms,
\begin{equation}\label{2.6}
(M+\frac{1}{2}U)_{uv}=0\ \Leftrightarrow\
e^{-M-\frac{1}{2}U}=\frac{1-k\,u}{(1+b^2v^2)^{3/2}}\
,\end{equation} using the boundary conditions (1.9) and (2.2). Now
the first of Maxwell's equations in (1.2) with $O(a^2)$--terms
neglected yields
\begin{equation}\label{2.7}
\frac{\partial\hat\phi _0}{\partial u}=0\ ,\end{equation}and
solving this using the boundary conditions (1.9) results in
\begin{equation}\label{2.8}
\hat\phi _0=\frac{b}{(1+b^2v^2)^{3/2}}\ ,\end{equation}neglecting
$O(a^2)$--terms. Now (1.4) and (1.5) simplify, neglecting
$O(a^2)$--terms, to
\begin{eqnarray}\label{2.9}
2U_{uu}&=&U_u^2-2\,U_u\,M_u\ ,\\
2\,U_{vv}&=&U_v^2-2\,U_v\,M_v+4\,\phi _0^2\
,\end{eqnarray}respectively. These are automatically satisfied by
$U, M, \phi _0$ given by (2.7), (2.8) and (2.10). The remaining
field equations, the second of (1.2) along with (1.6), will
determine the two remaining unknown functions $V$ (or equivalently
$A$) and $\hat\phi _2$.

To obtain $V$ we begin by eliminating $\phi _2$ from (1.6). We
start by writing (1.6) in the form
\begin{equation}\label{2.10}
2\,e^{-U}V_{uv}=-\left (e^{-U}\right )_uV_v-\left (e^{-U}\right
)_vV_u+4\,\hat\phi _0\,\hat\phi _2\ .\end{equation}Differentiating
this with respect to $v$ and multiplying by $\hat\phi _0$ yields
\begin{eqnarray}\label{2.11}
&&\hat\phi _0\left\{2\,e^{-U}V_{uvv}+3\,\left (e^{-U}\right
)_vV_{uv}+\left (e^{-U}\right )_uV_{vv}+\left (e^{-U}\right
)_{vv}V_u\right\}\nonumber\\
&=&-2\,\hat\phi _0^3V_u+\frac{\partial\hat\phi _0}{\partial
v}\left\{2\,e^{-U}V_{uv}+\left (e^{-U}\right )_uV_v+\left
(e^{-U}\right )_vV_u\right\}\ .\end{eqnarray}With $U$ and
$\hat\phi _0$ given by (2.7) and (2.10) respectively we find that
\begin{equation}\label{2.12}
\hat\phi _0\left (e^{-U}\right )_{vv}+2\,\hat\phi _0^3-\left
(e^{-U}\right )_v\frac{\partial\hat\phi _0}{\partial v}=0\
,\end{equation}and so the three terms in (\ref{2.11}) having $V_u$
as a factor disappear and (\ref{2.11}) becomes an equation for
$W=V_v$ given by
\begin{eqnarray}\label{2.13}
&&\left\{(1-k\,u)^2(1+b^2v^2)-b^2v^2\right\}W_{uv}=
-3\,b^2v\,\left\{(1-k\,u)^2-1\right\}\,W_u\nonumber\\&&
+k\,(1-k\,u)\,(1+b^2v^2)\,W_v+3\,k\,b^2v\,(1-k\,u)\,W\
.\end{eqnarray}To solve this equation it is useful to change the
dependent variable $W$ to a new variable ${\cal K}(u, v)$ (say) in
such a way that the resulting differential equation for ${\cal K}$
does not have an undifferentiated ${\cal K}$--term. Multiplying
(\ref{2.13}) by $v$ and writing
\begin{equation}\label{2.14}
W=\frac{{\cal K}_v}{1-k\,u}\ ,\end{equation}for some ${\cal K}$,
the equation (\ref{2.13}) implies
\begin{equation}\label{2.15}
\left\{(1-k\,u)\,(1+b^2v^2)-\frac{b^2v^2}{1-k\,u}\right\}\,W_u-k\,(1+b^2v^2)\,W
={\cal K}_u\ .\end{equation}Substituting for $W$ from (2.17) into
(2.18) results in
\begin{equation}\label{2.16}
\varphi\,{\cal K}_{uv}=\frac{b^2v^3k}{(1-k\,u)^3}\,{\cal
K}_v+{\cal K}_u\ ,\end{equation}where
\begin{equation}\label{1.2}
\varphi =v+b^2v^3\left (1-\frac{1}{(1-k\,u)^2}\right )\
.\end{equation}The important difference between this equation and
(2.16) is that in this equation the dependent variable ${\cal K}$
does not appear undifferentiated (whereas an undifferentiated $W$
appears in (2.16)). This makes it easier to solve (2.19) than
(2.16).\vskip 4truepc

\setcounter{equation}{0}
\section{Construction of a Candidate Solution ${\cal K}$}\indent

We look for the solution ${\cal K}$ of (2.19) satisfying the
following boundary conditions (which follow from (2.3), (2.4),
(2.17) and (2.18)) : When $u=0$ we must have ${\cal
K}_u=2\,a\,b\,v$\ and\ ${\cal K}_v=0$ and when $v=0$ we require
${\cal K}_u=0$ and ${\cal K}_v=2\,a\,b\,u$. We will henceforth
drop the factor $2\,a\,b$ for the moment and reinstate it at the
very end.

First we rewrite (2.19) to read
\begin{equation}\label{3.1}
v\,{\cal K}_{uv}-{\cal K}_u=b^2v^3\left
\{\frac{k}{(1-k\,u)^3}\,{\cal K}_{v}-\left
(1-\frac{1}{(1-k\,u)^2}\right )\,{\cal K}_{uv}\right \}\
.\end{equation} This, and the boundary conditions, suggest that
we look for a solution which is a power series in powers of $b^2$
of the form
\begin{equation}\label{3.2}
{\cal K}=u\,v+b^2{\cal K}^{(1)}+b^4{\cal K}^{(2)}+\dots\
.\end{equation}The first term here is reminiscent of a
corresponding term in the Bell--Szekeres \cite{BS} solution of the
Einstein--Maxwell vacuum field equations. The precise connection
with the Bell--Szekeres solution is mentioned in section 6 below.
Substitution of (3.2) into (3.1) and equating powers of $b^2$ on
both sides leads to a sequence of differential equations for
${\cal K}^{(1)}, {\cal K}^{(2)}, \dots$. The boundary conditions
above result in the first derivatives of these functions having to
vanish when $u=0$ and when $v=0$. This will fix the functions
uniquely up to a constant in each case. We can take this constant
to vanish because ultimately it can be removed by rescaling the
coordinates $x$ and $y$ in (1.1). The differential equations for
the coefficients in (3.2) read
\begin{eqnarray}\label{3.3}
v\,{\cal K}^{(1)}_{uv}-{\cal
K}^{(1)}_u&=&-v^3+\frac{v^3}{(1-k\,u)^3}\ ,\\
v\,{\cal K}^{(i)}_{uv}-{\cal
K}^{(i)}_u&=&v^3\left\{\frac{k}{(1-k\,u)^3}{\cal
K}^{(i-1)}_{v}-\left (1-\frac{1}{(1-k\,u)^2}\right )\,{\cal
K}^{(i-1)}_{uv}\right\}\ ,\end{eqnarray}for $i=2, 3, 4,\dots\
$.These equations are straightforward to solve subject to the
boundary conditions given above. The first few functions are given
by
\begin{eqnarray}\label{3.4}
{\cal
K}^{(1)}&=&\frac{v^3}{2}\left\{-u+\frac{1}{2\,k\,(1-k\,u)^2}\right\}\
,\\
{\cal
K}^{(2)}&=&\frac{3\,v^5}{8}\left\{u-\frac{1}{k\,(1-k\,u)^2}+\frac{3}{8\,k\,(1-k\,u)^4}
\right\}\ ,\\
{\cal
K}^{(3)}&=&\frac{5\,v^7}{16}\left\{-u+\frac{3}{2\,k\,(1-k\,u)^2}-\frac{9}{8\,k\,(1-k\,u)^4}
+\frac{5}{16\,k\,(1-k\,u)^6}\right\}\ .\nonumber\\
&&\end{eqnarray}Substituting into (3.2) we now have
\begin{eqnarray}\label{3.5}
{\cal
K}&=&u\,v-\frac{1}{2}b^2v^3u+\frac{3}{8}b^4v^5u-\frac{5}{16}b^6v^7u+\dots\
\nonumber\\
&&-\frac{1}{2}\frac{b^2v^3}{k}\,\left (-\frac{1}{2}\,\chi\right
)+\frac{3}{8}\frac{b^4v^5}{k}\,\left (-\frac{1}{2}\,\chi
+\frac{3}{8}\,\chi ^2\right
)\nonumber\\
&&-\frac{5}{16}\frac{b^6v^7}{k}\,\left (-\frac{1}{2}\,\chi
+\frac{3}{8}\,\chi ^2-\frac{5}{16}\,\chi ^3\right )+\dots\
,\end{eqnarray}where the variable
\begin{equation}\label{3.5'}
\chi =\frac{1}{(1-k\,u)^{2}}-1\ ,\end{equation}has been introduced
for convenience. This variable (which vanishes when $u=0$) will
appear frequently in the sequel. Remembering that we are here
constructing \emph{a candidate} exact solution of (2.19)
satisfying the boundary conditions (the candidate will be verified
to be an exact solution in the next section), the form of (3.8)
suggests we should write
\begin{eqnarray}\label{3.6}
{\cal K}&=&\frac{u\,v}{\sqrt{1+b^2v^2}}\nonumber\\
&&-\frac{1}{2}\frac{b^2v^3}{k}\,\left (1-\frac{1}{2}\,\chi\right
)+\frac{3}{8}\frac{b^4v^5}{k}\,\left (1-\frac{1}{2}\,\chi
+\frac{3}{8}\,\chi ^2\right
)\nonumber\\
&&-\frac{5}{16}\frac{b^6v^7}{k}\,\left (1-\frac{1}{2}\,\chi
+\frac{3}{8}\,\chi ^2-\frac{5}{16}\,\chi ^3\right )+\dots\
\nonumber\\
&&+\left
[\frac{1}{2}\frac{b^2v^3}{k}-\frac{3}{8}\frac{b^4v^5}{k}+\frac{5}{16}\frac{b^6v^7}{k}-
\dots\ \right ]\ .\end{eqnarray}The final series in square
brackets here can tentatively be written
\begin{equation}\label{3.7}
-\frac{v}{k}\,\left\{\frac{1}{\sqrt{1+b^2v^2}}-1\right\}\
.\end{equation}Putting this into (3.10) results in
\begin{equation}\label{3.8}
{\cal
K}=-\frac{v\,(1-k\,u)}{k\,\sqrt{1+b^2v^2}}+\frac{v}{k}\,{\cal F}\
,\end{equation}with
\begin{eqnarray}\label{3.12}
{\cal F}&=&1-\frac{1}{2}b^2v^2\,\left (1-\frac{1}{2}\,\chi\right
)+\frac{3}{8}b^4v^4\,\left (1-\frac{1}{2}\,\chi +\frac{3}{8}\,\chi
^2\right
)\nonumber\\
&&-\frac{5}{16}b^6v^6\,\left (1-\frac{1}{2}\,\chi
+\frac{3}{8}\,\chi ^2-\frac{5}{16}\,\chi ^3\right )+\dots\
.\end{eqnarray}From this we calculate that
\begin{eqnarray}\label{3.13}
{\cal F}+2\,(\chi +1)\,\frac{\partial {\cal
F}}{\partial\chi}&=&1+3\,\left (\frac{1}{2}\right )^2b^2v^2\chi
+5\,\left (\frac{3}{8}\right )^2\,b^4v^4\chi ^2\nonumber\\
&&+7\,\left (\frac{5}{16}\right )^2\,b^6v^6\,\chi ^3+9\,\left
(\frac{35}{128}\right )^2\,b^8v^8\chi ^4+\dots\ .\end{eqnarray}We
can rewrite this in the form
\begin{equation}\label{3.14}
{\cal F}+2\,(\chi +1)\,\frac{\partial {\cal
F}}{\partial\chi}=\frac{\partial}{\partial v}(v\,{\cal G})\
,\end{equation}with
\begin{eqnarray}\label{3.15}
{\cal G}&=&1+\left (\frac{1}{2}\right )^2b^2v^2\chi
+\left (\frac{3}{8}\right )^2\,b^4v^4\chi ^2\nonumber\\
&+&\left (\frac{5}{16}\right )^2\,b^6v^6\,\chi ^3+\left
(\frac{35}{128}\right )^2\,b^8v^8\chi ^4+\dots\
.\end{eqnarray}This suggests that we should write
\begin{equation}\label{3.16}
{\cal G}=\sum_{n=0}^{\infty}\left
[\frac{(2\,n)!}{2^{2\,n}(n!)^2}\right ]^2(y\,\chi
)^n=\frac{2}{\pi}\,K(y\,\chi )\ ,\end{equation}where $y=b^2v^2$
and $K$ is the complete elliptic integral of the first kind (see
Eq.(A.1)) with argument $y\,\chi$. Substituting this into (3.15)
we can then integrate with respect to $\chi$ to obtain
\begin{equation}\label{3.17}
{\cal F}=\frac{2}{\pi}\,\Pi (-y, y\,\chi )+c_0(y)\ ,\end{equation}
where $\Pi$ is the complete elliptic integral of the third kind
(see Eq.(A.3)) and $c_0$ is a function of integration. The
boundary conditions are satisfied provided ${\cal F}=(1+y)^{-1/2}$
when $\chi =0\ \Leftrightarrow u=0$. This is true of (3.18)
provided the function of integration $c_0(y)=0$. Now (3.12)
multiplied by the factor $2\,a\,b$, and with ${\cal F}$ given by
(3.18) with $c_0=0$, is a candidate exact solution of (2.19)
satisfying the required boundary conditions. \vskip 4truepc
\setcounter{equation}{0}
\section{Verifying the Candidate Solution ${\cal K}$}\indent

The candidate solution constructed in the previous section reads:
\begin{equation}\label{4.1}
{\cal
K}=-\frac{2\,a\,b\,v\,(1-k\,u)}{k\,\sqrt{1+b^2v^2}}+\frac{4\,a\,b}{\pi\,k}\,
v\,\Pi (-b^2v^2, b^2v^2\chi )\ .\end{equation}It is easy to check
that the first term satisfies the differential equation (2.19).
Hence we wish to demonstrate that
\begin{equation}\label{4.2}
\tilde {\cal K}=v\,\Pi (-b^2v^2, b^2v^2\chi )\
,\end{equation}satisfies (2.19). Writing $n=-b^2v^2$ and
$m=b^2v^2\chi$ and substituting (4.2) into (2.19) results in $\Pi
(n, m)$ having to satisfy the differential equation
\begin{equation}\label{4.3}
2\,n\,(m-1)\,\frac{\partial ^2\Pi}{\partial n\partial
m}+2\,m\,(m-1)\,\frac{\partial ^2\Pi}{\partial
m^2}+n\,\frac{\partial\Pi}{\partial
n}+2\,(2\,m-1)\,\frac{\partial\Pi}{\partial m}+\frac{1}{2}\Pi =0\
.\end{equation}Using the formula (A.6) for the partial derivative
of $\Pi$ with respect to $m$ this can be rewritten in the form
\begin{eqnarray}\label{4.4}
n\,(n-1)\,\frac{\partial\Pi}{\partial
n}&+&(-3\,m^2+m+4\,m\,n-2\,n)\frac{\partial\Pi}{\partial
m}+\frac{1}{2}\,(-3\,m+n+2)\,\Pi\nonumber\\
&+&\frac{(1-3\,m)}{2\,(m-1)}\,E-\frac{1}{2}K=0\
,\end{eqnarray}where $K$ and $E$ are the complete elliptic
integrals of the first and second kind respectively (see
Appendix). The formulas giving the partial derivatives of $\Pi$
with respect to $m$ and $n$ in terms of the complete elliptic
integrals are given in (A.6) and (A.7) respectively. When these
substitutions are made in (4.4) it follows that (4.4) is an
identity. Hence the candidate solution (4.1) of (2.19) is indeed
an exact solution of (2.19).
\vskip 4truepc
\setcounter{equation}{0}
\section{Perturbed Fields}\indent

In order to demonstrate explicitly that the perturbed field we are
constructing here contains backscattered gravitational radiation
we must use the function ${\cal K}$ we have obtained to calculate
the leading terms (for small $a$) in $\Psi _0$ and $\Psi _4$ given
by (1.13) and (1.17) and see that they are non--vanishing. This we
can now do because with (2.5) and (2.17), with $W=V_v$, we have
\begin{equation}\label{5.1}
V=2\,A=\frac{{\cal K}}{(1-k\,u)}\ ,\end{equation} neglecting
$O(a^2)$--terms, with ${\cal K}$ given by (4.1). In integrating
(2.17) with respect to $v$ a function of $u$ of integration has
been put equal to zero on account of the fact that by (2.2) $V$
must vanish when $v=0$ and we have already ensured that ${\cal K}$
vanishes when $v=0$. Since $V$ is small of order $a$ we can enter
the `background' expressions (2.7) and (2.8) for $U$ and $M$ in
(1.13). This initially results in
\begin{equation}\label{5.2}
\Psi
_0=-\frac{1}{2}V_{vv}-\frac{3\,b^2v}{2\,(1+b^2v^2)}\,V_v+\frac{3}{4}\,U_v\,V_v\
.\end{equation}Introducing again $n=-b^2v^2$ and $m=b^2v^2\chi$ we
obtain from (2.7) the expression
\begin{equation}\label{5.3}
U_v=-\frac{2\,(n-m)}{v\,(n-1)\,(m-1)}\ .\end{equation}By (5.1)
with ${\cal K}$ given by (4.1) we obtain, using (A.6) and (A.7),
\begin{equation}\label{5.4}
V_v=-\frac{2\,a\,b}{k\,(1-n)^{3/2}}+\frac{4\,a\,b}{\pi\,k\,(1-k\,u)\,(n-1)}
\left\{-\Pi +K+\frac{1}{m-1}E\right\}\ ,\end{equation}and
\begin{equation}\label{5.5}
\frac{1}{2}V_{vv}+\frac{3\,b^2v}{2\,(1+b^2v^2)}\,V_v=-\frac{2\,a\,b}{\pi\,k\,v\,(1-k\,u)
\,(n-1)\,(m-1)}\left\{K+\frac{(m+1)}{(m-1)}\,E\right\}\
,\end{equation}with $\Pi$ a function of $n, m$ and $K, E$
functions of $m$. Thus $\Psi _0$ in (5.2) has the non--zero value
\begin{eqnarray}\label{5.6}
\Psi _0&=&\frac{2\,a\,b}{\pi\,k\,v\,(1-k\,u)\,(n-1)^2(m-1)}\,\{
3\,(n-m)\,\Pi-(2\,n-3\,m+1)\,K\nonumber\\
&&-\frac{(2\,n-2\,m-n\,m+1)}{(m-1)}\,E\}-\frac{3\,a\,b\,(n-m)}{k\,v\,(1-n)^{5/2}(m-1)}\
,\end{eqnarray}which indicates that following the collision of the
light--like shell, accompanied by the small amplitude
electromagnetic waves labelled by $a$, with the electromagnetic
waves labelled by $b$ there exist backscattered gravitational
waves having propagation direction in space--time $\partial
/\partial u$. In similar fashion we find that
\begin{equation}\label{5.7}
V_u=-\frac{4\,a\,b\,v}{\pi\,(1-k\,u)^2(m-1)}\,E\
,\end{equation}and
\begin{equation}\label{5.8}
\Psi _4=-\frac{2\,a\,b\,k\,v}{\pi\,\chi\,(1-k\,u)^3(m-1)}\left\{
K-\frac{(2\,m-1)}{(m-1)}\,E\right\}\ .\end{equation}The
non--vanishing of this quantity indicates the existence of
backscattered gravitational radiation after the collision having
propagation direction in space--time $\partial /\partial v$. Since
$V$ is small of first order and, neglecting $O(a^2)$--terms, $U$
is given by (2.7), we see that $\Psi _2$ in (1.15) has its
background value in the linear approximation given by
\begin{equation}\label{5.9}
\Psi _2=\frac{k\,n}{v\,(1-k\,u)^3(m-1)^2}\ .\end{equation}

To discover what type of physical signal has as its history in
space--time the boundary $u>0, v=0$ or $v>0, u=0$ of the
interaction region of the space--time we carry out the calculation
above replacing $u$ with $u_+=u\,\vartheta (u)$ and $v$ with
$v_+=v\,\vartheta (v)$ where $\vartheta$ is the Heaviside
step--function. We find that in addition to the backscattered
gravitational radiation found above there are Dirac
delta--function terms in $\Psi _0$ and $\Psi _4$ given by
\begin{equation}\label{5.10}
{}^{\delta}\Psi _0=-\frac{a\,b\,u_+}{(1-k\,u_+)}\,\delta (v)\
,\qquad {}^{\delta}\Psi _4=-a\,b\,v_+\,\delta (u)\ ,\end{equation}
indicating that the boundaries of the interaction region are the
histories of impulsive gravitational waves. We note that the Ricci
tensor possesses a delta function term,
${}^{\delta}R_{ij}=-2\,k\,\delta (u)\,u_{,i}\,u_{,j}$, reflecting
the presence of the light--like shell (labelled by $k$) having
history $u=0$.

The Maxwell field after the collision has two radiative components
conveniently described by $\hat\phi _0$ and $\hat\phi _2$. In the
approximation in which $O(a^2)$--terms are neglected we have
already seen that $\hat\phi _0$ is given by (2.10). With ${\cal
K}$ and thus $V$ already known in this approximation we turn to
the second of the Maxwell equations (1.2) to find $\hat\phi _2$.
Neglecting $O(a^2)$--terms this equation reads
\begin{equation}\label{5.11}
\frac{\partial\hat\phi _2}{\partial
v}=-\frac{b}{2\,(1+b^2v^2)^{3/2}}\,\frac{\partial}{\partial
u}\left (\frac{{\cal K}}{(1-k\,u)}\right )\ .\end{equation}Using
(4.1) and (A.3) with, for convenience, the substitution $b^2v^2=y$
this simplifies to
\begin{equation}\label{5.12}
\frac{\partial\hat\phi _2}{\partial
y}=-\frac{a}{\pi\,(1-k\,u)^2(1+y)^{3/2}}\int_{0}^{\pi
/2}\frac{d\theta}{(1-y\,\chi\,\sin ^2\theta )^{3/2}}\
.\end{equation}Noting that
\begin{equation}\label{5.13}
\frac{d}{dy}\left (\frac{1-\chi\,(1+2\,y)\,\sin
^2\theta}{\sqrt{1+y}\,\sqrt{1-y\,\chi\,\sin ^2\theta}}\right )=
-\frac{(1+\chi\,\sin ^2\theta )^2}{2\,(1+y)^{3/2}(1-y\,\chi\,\sin
^2\theta )^{3/2}}\ ,\end{equation}we can integrate (5.12) to
obtain
\begin{equation}\label{5.14}
\hat\phi _2=\frac{2\,a}{\pi\,(1-k\,u)^2\sqrt{1+y}}\,\int_{0}^{\pi
/2}\frac{(1-\chi\,(1+2\,y)\,\sin ^2\theta
)\,d\theta}{(1+\chi\,\sin ^2\theta )^2\sqrt{1-y\,\chi\,\sin
^2\theta}}\ .\end{equation}There is no need for a function of
$\chi$ of integration since evaluating this at $v=0$ (which
corresponds to $y=0$) yields the correct boundary value of
$\hat\phi _2=a\,(1-k\,u)$ (see Eq.(2.2)). Rearranging the
integrand in (5.14) permits us to write
\begin{eqnarray}\label{5.15}
\hat\phi
_2&=&-\frac{4\,a\,\chi\,\sqrt{1+y}}{\pi\,(1-k\,u)^2}\int_{0}^{\pi
/2}\frac{\sin ^2\theta\,d\theta}{(1+\chi\,\sin ^2\theta
)^2\sqrt{1-y\,\chi\,\sin ^2\theta}}\nonumber\\
&&+\frac{2\,a}{\pi\,(1-k\,u)^2\sqrt{1+y}}\,\Pi (-\chi , y\,\chi )\
,\end{eqnarray}with $\Pi (n, m)$ given by (A.3). Making use of
(A.7) we have
\begin{eqnarray}\label{5.16}
\int_{0}^{\pi /2}\frac{\sin ^2\theta\,d\theta}{(1+\chi\,\sin
^2\theta )^2\sqrt{1-y\,\chi\,\sin ^2\theta}}&=&\frac{(\chi
-y)\,\Pi (-\chi , y\,\chi )}{2\,\chi\,(\chi
+1)\,(y+1)}+\frac{K(y\,\chi )}{2\,\chi\,(\chi
+1)}\nonumber\\&&-\frac{E(y\,\chi )}{2\,\chi\,(\chi +1)\,(y+1)}\
,\end{eqnarray}and this simplifies (5.15) to read
\begin{equation}\label{5.17}
\hat\phi _2=\frac{2\,a}{\pi}\,\sqrt{1+y}\,\left\{\Pi (-\chi ,
y\,\chi )-K(y\,\chi )\right\}+\frac{2\,a}{\pi}\,\frac{E(y\,\chi
)}{\sqrt{1+y}}\ .\end{equation}Due to the presence of $\hat\phi
_0$ and $\hat\phi _2$ after the collision, two systems of
backscattered electromagnetic radiation exist having propagation
directions $\partial /\partial u$ and $\partial /\partial v$ in
the interaction space--time. \vskip 4truepc
\setcounter{equation}{0}
\section{Discussion}\indent

Neglecting $O(a^2)$--terms we have obtained explicit expressions
for the functions $U, M, V$ in the line--element (1.1). These are
given respectively by (2.7), (2.8) and by (5.1) with (4.1). We
also have explicit expressions for the Maxwell field described by
the functions $\hat\phi _0$ and $\hat\phi _2$. These functions are
found in (2.10) and (5.17) respectively.

The approximate solution, for small $a$, of the Einstein--Maxwell
field equations described above has a limit $k\rightarrow 0$
corresponding to the removal of the light--like shell. In this
limit (4.1) gives ${\cal K}\rightarrow 2\,a\,b\,u\,v$ while $\Psi
_0\rightarrow 0$ by (5.6), $\Psi _4\rightarrow 0$ by (5.8) and
$\Psi _2\rightarrow 0$ by (5.9) so that the backscattered
gravitational waves disappear, but the impulsive gravitational
waves (5.10) remain with $k=0$ in ${}^{\delta}\Psi _0$. All of
this corresponds to the important Bell--Szekeres \cite{BS}
solution giving the exact conformally flat space--time following
the collision of the two electromagnetic shock waves labelled by
$a$ and $b$. As a converse to the problem considered in this paper
one might ask whether the introduction of a light--like shell
accompanying one of these waves would produce backscattered
gravitational radiation after collision. It is sufficient to
answer this question in the affirmative by assuming small $a$ and
small $k$, where $k$ is proportional to the energy density of the
light--like shell measured by specified observers \cite{BH2}, and
to specialize $\Psi _0$ and $\Psi _4$, given by (5.6) and (5.8)
respectively, to this case and see that they are non--zero. We
find that
\begin{eqnarray}\label{6.1}
\Psi _0&=&a\,b^3k\,u^2v\,\left\{\frac{9\,(2+7\,b^2v^2+4\,b^4v^4)}{
(1+b^2v^2)^{5/2}}+\frac{7+7\,b^2v^2+3\,b^4v^4}{2\,(1+b^2v^2)^2}\right\}\ ,\\
\Psi _4&=&\frac{3}{2}\,a\,b^3k\,v^3\ ,\end{eqnarray}neglecting
$O(k^2)$--terms. In addition (5.9) becomes, for small $k$,
\begin{equation}\label{6.2}
\Psi _2=-b^2k\,v+O(k^2)\ ,\end{equation}and these last three
formulas confirm the limits mentioned above.

\noindent
\section*{Acknowledgment}\noindent
We thank Julien Garaud for helpful comments and Universit\'e de
Tours and Ambassade de France en Irlande for financial support.

\vskip 4truepc
\appendix
\section{Formulas for Complete Elliptic Integrals} \setcounter{equation}{0}
The complete elliptic integrals of the first, second and third
kinds are given respectively by \cite{GR}
\begin{eqnarray}\label{2.10}
K&=&\int_{0}^{\pi /2}\frac{d\theta}{\sqrt{1-m\,\sin ^2\theta}}=K(m)\ ,\\
E&=&\int_{0}^{\pi /2}\sqrt{1-m\,\sin ^2\theta}\,d\theta =E(m)\ ,\\
\Pi&=&\int_{0}^{\pi /2}\frac{d\theta}{(1-n\,\sin ^2\theta
)\,\sqrt{1-m\,\sin ^2\theta}}=\Pi (n, m)\ ,\end{eqnarray}where $m,
n$ are real constants. The well--known formulas for their
derivatives are
\begin{eqnarray}\label{A-1}
\frac{dK}{dm}&=&-\frac{1}{2\,m}\,\left (K+\frac{1}{m-1}\,E\right
)\ ,\\
\frac{dE}{dm}&=&\frac{1}{2\,m}(E-K)\ ,\\
\frac{\partial\Pi}{\partial m}&=&\frac{1}{2\,(n-m)}\left\{\Pi
+\frac{1}{m-1}\,E\right\}\ ,\\
\frac{\partial\Pi}{\partial
n}&=&-\frac{1}{2\,(n-m)\,(n-1)}\left\{\frac{(n^2-m)}{n}\,\Pi
-\frac{(n-m)}{n}\,K+E\right\}\ .\end{eqnarray}

\end{document}